\documentclass{osa-article} 
\journal{oe}
\usepackage{cite}
\usepackage{amsmath,amssymb,amsfonts}
\usepackage{graphicx}
\usepackage{textcomp}
\usepackage{xcolor}
\usepackage{verbatim}
\usepackage{float}
\usepackage{bm}
\usepackage{subfigure}

\usepackage{algorithm}
\usepackage{algorithmicx}
\usepackage{algpseudocode}

\def\BibTeX{{\rm B\kern-.05em{\sc i\kern-.025em b}\kern-.08em
    T\kern-.1667em\lower.7ex\hbox{E}\kern-.125emX}}
\begin{document}
\title{Power Efficient LED Placement Algorithm for Indoor Visible Light Communication}

\author{Yang Yang,\authormark{1,*} Zhiyu Zhu,\authormark{1,} Caili Guo,\authormark{2} and Chunyan Feng\authormark{1,2}}

\address{\authormark{1}Beijing Key Laboratory of Network System Architecture and Convergence, School of Information and Communication Engineering, Beijing University of Posts and Telecommunications, Beijing 100876, China\\
\authormark{2}Beijing Laboratory of Advanced Information Networks, School of Information and Communication Engineering, Beijing University of Posts and Telecommunications, Beijing 100876, China}

\email{\authormark{*}young0607@bupt.edu.cn} 

\vspace{0cm}
\begin{abstract}
This paper proposes a novel power-efficient light-emitting diode (LED) placement algorithm for indoor visible light communication (VLC). In the considered model, the LEDs can be designedly placed for high power efficiency while satisfying the indoor communication and illumination requirements. This design problem is formulated as a power minimization problem under both communication and illumination level constraints. Due to the interactions among LEDs and the illumination uniformity constraint, the formulated problem is complex and non-convex. To solve the problem, we first transform the complex uniformity constraint into a series of linear constraints. Then an iterative algorithm is proposed to decouple the interactions among LEDs and transforms the original problem into a series of convex sub-problems. Then, we use Lagrange dual method to solve the sub-problem and obtain a convergent solution of the original problem. Simulation results show that the proposed LED placement algorithm can harvest $22.86\% $ power consumption gain when compared with the baseline scheme with centrally placed LEDs.
\end{abstract}

%

\vspace{-0.cm}
\section{Introduction}
\label{sec:intro}
Artificial illumination consumes approximately $20\%$ of the world's electricity \cite{World2020}. Under this background, light emitting diode (LED) has gained increasing market rate due to its high power efficiency and low cost. Moreover, a promising short range wireless access technology based on LEDs, termed as visible light communication (VLC), has also gained increasing attentions \cite{YangAnEnhanced}.
Compared with conventional radio-frequency based access technologies, VLC has high power efficiency due to the use of visible light. In addition, It also has the advantage of abundant bandwidth resource, inherent high security, flexibility and rapid deployment time \cite{wang2019privacy}.

Besides these promising advantages, VLC also has some key design issues that should be carefully tackled, and the indoor LED arrangement problem is one of them. In particular, VLC involves both communication and illumination. That means besides the communication requirement of users, the illumination requirement of users should also be considered for a VLC system. Besides, since indoor illumination typically requires a number of closely located LEDs for sufficient illuminance level, the illumination and communication interactions of LEDs need to be considered in indoor LED arrangement. Hereafter, we use the interaction of LEDs to indicate the fact that the signal from each LED can be constructively added at the receiver from the perspective of illumination while that can also be the interference of each other from the perspective of communication. In addition, the illuminance at the receiver plane should be uniform, which is preferred for indoor illumination engineering \cite{Wang2012Performance}. All these factors need to be considered in indoor LED arrangement designs, which make the problem a challenging design issue.

A plethora of studies on indoor LED arrangement have been proposed, which can be classified into two types \cite{varma2018optimum}. One type of approach is by appropriately allocating power to LEDs \cite{Wang2012Performance,varma2018optimum,varma2017power,mushfique2020optimization}. In \cite{varma2018optimum}, the uniform illuminance problem was formulated as a convex problem, which was then solved by quadratic programming to optimize the power of each LED. In \cite{varma2017power}, a heuristic power allocation scheme was proposed for a random LED array to obtain uniform irradiance on the projection surface. In both \cite{varma2018optimum} and \cite{varma2017power}, a quality factor that measures the illuminance performance was used to quantify the illumination performance. However, the communication performance of the system was not considered in \cite{varma2018optimum} and \cite{varma2017power}. Note that satisfactory illumination does not necessarily mean that the communication requirement of users are satisfied \cite{liu2014cellular}.

The other type of approach is by optimizing the placement of LEDs \cite{stefan2013analysis,liu2014cellular,ali2016energy,liu2017optimization}.
In \cite{liu2014cellular}, a genetic LED placement algorithm was used to maximize the coverage of illumination and communication of an indoor VLC system.
In \cite{stefan2013analysis}, the optimal LED placement for the maximum average area spectral efficiency was investigated. Due the complexity of the problem, computer simulations are used to obtain the optimal solutions \cite{stefan2013analysis}.
The work in \cite{ali2016energy} only considered the illumination constraint, the power efficient LED placement problem was formulated as a linear program problem and was thus solved efficiently.
While interesting, in \cite{Wang2012Performance,varma2018optimum,varma2017power,mushfique2020optimization,stefan2013analysis,liu2014cellular,ali2016energy,liu2017optimization}, the interactions between LEDs are not considered, which can significantly affect the signal-to-noise ratio (SNR) at the receiver, thus further altering the optimal indoor LED placement.

The main contribution of this paper is a power efficient LED placement algorithm that takes both the illumination and communication interactions of different LEDs into consideration. First, the system model of communication and illumination of indoor VLC is studied, based on which a power efficient LED placement optimization framework is formulated under the constrains of communication, illumination level and the indoor illumination uniformity constraint. Due to the complexity of the illumination uniformity constraint, we first transform it into a series of linear constrains. But, still, the problem after the transformation is non-convex due to the close interdependence of each optimization variables. To decouple the interdependence of the optimization variables, we then propose an iterative algorithm that transforms the original problem into a series of convex, independent sub-problems. The Lagrange dual method is used to solve the sub-problems and the solutions of each sub-problem will be iteratively plugged into the next sub-problem until a convergent solution is obtained. Simulation results show that the proposed LED placement algorithm can harvest 22.86$\% $ power consumption gain when compared with the baseline
with LEDs at the center of each sub-area.

The remainder of this paper is organized as follows. The system model is depicted in Section \ref{sys model}, where the problem is formulated. In Section \ref{proposed}, the proposed power efficient LED placement algorithm is detailed, and in Section  \ref{results} numerical results are presented and analyzed. Finally, Section \ref{Conclusion} draws some important conclusions.
\vspace{-0.cm}
\section{System Model}\label{sys model}
\begin{figure}[htbp]
\vspace{-0.5cm}
\centering
\includegraphics[height=5cm,trim=1 1 0 0,clip]{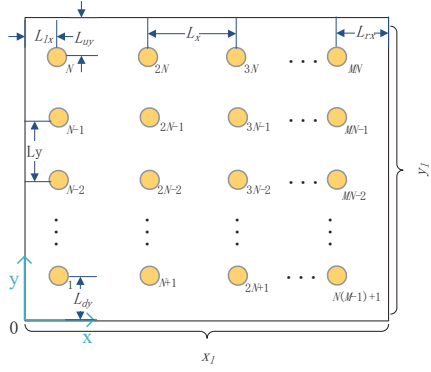}
\caption{The top view of the ceiling. Without loss of generality, each LED is marked by an index.}\label{scenario}
\vspace{-0.5cm}
\end{figure}

We consider an indoor VLC system with an equally spaced, LED array set ${\cal K}$ of $K$ LEDs. Fig. \ref{scenario} shows the top view of LEDs' distribution in a $x_l \times y_l$ room. The LEDs provide downlink transmission and illumination for the receivers, simultaneously. A set ${\cal U}$ of $U$ receivers is assumed to be uniformly distributed at the receiver plane so that the whole indoor illumination and communication performance can be tested by the performance of $U$ receivers.
\vspace{-0.1cm}

In the considered model, the multipath propagation resulted from reflections and refractions is neglected, and only line of sight (LOS) channel model is considered \cite{moreno2008modeling}. Given an LED $i\in{\cal K}$ and a receiver $j\in{\cal U}$, the channel gain between LED $i$ and receiver $j$ can be represented as \cite{yang2019relay}:
\begin{equation}\label{E-channel gain}
{h_{ij}} \!= \!\left\{\!\! \begin{aligned}
\frac{{(m \!+ \!1)A}}{{2\pi {d_{ij}}^2}}g\left( {{\psi _{ij}}} \!\right){\cos\! ^m}\left(\! {{\phi _{ij}}}\! \right)\cos\! \left(\! {{\psi _{ij}}} \right),0\! \le \!{\psi _{ij}} \!\le\! {\Psi _c},\!\!\\
0,{\kern 9em}  {\psi _{ij}}\! >\! {\Psi _c},\!\!
\end{aligned} \!\!\right.
\end{equation}
where $A$ is the detector area and ${d_{ij}}$ is the distance between LED $i$ and receiver $j$. In addition, $m = -\ln 2/\ln \left( {\cos {\Phi _{1/2}}} \right)$ is the Lambert order with ${\Phi _{1/2}}$  being the transmitter semi-angle. ${\psi _{ij}}$  is the angle of incidence, ${{\phi }_{ij}}$  is the angle of irradiance, ${\Psi _c}$ is the receiver field of vision (FOV) semi-angle, and $g\left( {{\psi }_{ij}} \right)$ is the gain of optical concentrator, which can be defined as:
\begin{equation}
  g\left( {{\psi _{ij}}} \right)=\left\{ \begin{array}{r}
\frac{{{n_r}^2}}{{{{\sin }^2}{\Psi _c}}},0 \le {\psi _{ij}} \le {\Psi _c},\\
0,{\kern 3em}{\psi _{ij}} > {\Psi _c},
\end{array} \right.
\end{equation}
where ${n_r}$ represents the refractive index. The height difference between the ceiling and the horizontal plane is fixed to $H$. Then the distance between LED $i$ and receivers $j$ is ${d_{ij}} = \sqrt {{{\left( {{x_i} - {x_j}} \right)}^2} + {{\left( {{y_i} - {y_j}} \right)}^2} + {H^2}}$, where $\left( {{x_j},{y_j}} \right), j = 1,2,...U$ denote the 2-dimension position of the receivers at the receiver plane and $\left( {{x_i},{y_i}} \right), i = 1,2,...K$ denote the 2-dimension position of the LEDs at the transmitter plane.

For receiver $j$ located at $\left( {{x}_{j}},{{y}_{j}} \right)$, the channel capacity ${{c}_{ij}}$ of the VLC link between LED $i$ and receiver $j$ can be given by \cite{wang2013tight}:
\begin{equation}\label{E-capacity}
{c_{ij}} = \frac{1}{2}{\log _2}\left( {1 + \frac{e}{{2\pi }}\frac{{{{\left( {\xi {P_i}{h_{ij}}} \right)}^2}}}{{{\sigma _\omega }^2 + \sum\nolimits_{n = 1,n \ne i}^K {(\xi {P_n}{h_{nj}}} {)^2}}}} \right) ,
\end{equation}
where $\xi$ is the illumination target, ${{P}_{i}}$ is the optical power of LED $i$, ${{\sigma }_{\omega }}$ is the standard deviation of the additive white Gaussian noise and $\sum\nolimits_{n = 1,n \ne i}^K {(\xi {P_n}{h_{nj}}} {)^2}$ is sum interferences from other LEDs. The ambient light noise is modeled as Gaussian noise \cite{komine2004fundamental}.

The illumination is another constraint that should be considered. In particular, both the illumination level and the indoor illumination uniformity should be considered. In this work, we use illuminance to indicate the illumination level, and the illuminance at the receiver $j$ from LED $i$ is proportional to ${\eta _{ij}} = \xi {P_i}{h_{ij}}$ \cite{din2014energy}.

As for the indoor illumination uniformity level, we use coefficient of variation of root mean square error CV(RMSE) to evaluate \cite{su2012designing}, which is defined as:
\begin{equation}\label{cvrmse}
   \setlength{\abovedisplayskip}{1pt}
   \setlength{\belowdisplayskip}{1pt}
{\rm{CV}(RMSE)} = \frac{{\rm RMSE}}{{{\eta _{avg}}}},
\end{equation}
where $\eta _{avg}$ is the average illumination, and it is given by:
\begin{equation}\label{etaavg}
   \setlength{\abovedisplayskip}{1pt}
   \setlength{\belowdisplayskip}{1pt}
{\eta _{avg}} = \frac{1}{U}\sum\limits_{i = 1}^K {\sum\limits_{j = 1}^U {\xi {P_i}{h_{ij}}} }.
\end{equation}
The illumination root mean square error at the receiver plane can be represented as:
\begin{equation}\label{rmse}
   \setlength{\abovedisplayskip}{1pt}
   \setlength{\belowdisplayskip}{1pt}
{\rm RMSE} = \sqrt {\frac{1}{U}\sum\limits_{i = 1}^K {\sum\limits_{j = 1}^U {{{\left( {\xi {P_i}{h_{ij}} - {\eta _{avg}}} \right)}^{\rm{2}}}} } }.
\end{equation}
\vspace{-0.2cm}
\subsection{Problem formulation}\label{profml}
The target of this work is to find the optimal LED placement that minimizes the LED power consumption under the data rate, illumination level and illumination uniformity constraints. In the considered model, we assume that each receiver will be serviced by the LED that provides the strongest signal strength, and the signals from other LEDs will be deemed as interference. At the receiver plane, we assume that each user has a data
rate constraint $c_{th}$ and a illumination level constraint $\eta_{th}$. To satisfy the data rate constraint of receiver $j$, the required power of LED $i$ can be derived from (\ref{E-capacity}) as:
\begin{equation}\label{E-PfromC}
{P_{ij}} = \frac{{\sqrt {{\sigma _\omega }^2 + \sum\nolimits_{n = 1,n \ne i}^K {(\xi {P_n}{h_{nj}}} {)^2}} \sqrt {\frac{{2\pi }}{e}\left( {{2^{2{c_{th}}}} - 1} \right)} }}{{\xi {h_{ij}}}}.
\end{equation}
For each LED, once the receiver with the maximum power requirement is satisfied, other receivers' constraints can be satisfied. Therefore, the minimum transmit power of LED $i$ is given by ${P_{i,\min }} = \max \{ {P_{ij}}\} ,\forall j \in {{\cal U}_i}$, where ${{\mathcal{U}}_{i}}$ represents the set of receivers associated with LED $i$. The received illuminance of receiver $j$ is given by ${\eta _{j}} = \sum\nolimits_{i = 1}^K {\xi {P_i}{h_{ij}}}$. To satisfy the illumination level constraint, the illuminance that the receiver $j$ receives should not be less than  ${\eta _{th}}$, i.e., $\sum\nolimits_{i = 1}^K {\xi {P_i}{h_{ij}}}  \ge {\eta _{th}}$. Thus, the illuminance that the receiver $j$ actually receives from its associated LED $i$ should satisfy:
\begin{equation}
\xi{P_i}{h_{ij}}\ge{{ I}_{ij}},
\end{equation}
where ${{ I}_{ij}} = {\eta _{th}} - \sum\nolimits_{n = 1,n \ne i}^K {\xi {P_n}{h_{nj}}}$, and $\sum\nolimits_{n = 1,n \ne i}^K {\xi {P_n}{h_{nj}}}$ represents the sum illuminance that receiver $j$ obtains from other LEDs.

Let the illumination uniformity constraint be $U_{th}$. Then, the optimization problem is formulated as:
\begin{align}\label{opti problem}
&\mathop {\min }\limits_{{x_i},{y_i}}{\mkern 1mu}\sum\limits_{i = 1}^K {{P_i}} \\
&{\rm s.t.}{\kern 1em}\xi{P_i}{h_{ij}}\ge{I_{ij}},\forall i\in{\cal K},\forall j\in{{\cal U}_i} \label{E-opti problemA} \tag{\theequation a}, \\
& {\kern 2em} {P_i} \ge {P_{i,\min }},\forall i \in {\cal K}, \label{E-opti problemB} \tag{\theequation b}\\
& {\kern 2em}{\rm{CV}(RMSE)} \le U_{th} \label{E-opti problemC} \tag{\theequation c},\\
& {\kern 2em}{x_i} = {x_{n,}}\forall i,n \in {\cal K},\left\lfloor {i/N} \right\rfloor  = \left\lfloor {n/N}\right\rfloor \label{E-opti problemD} \tag{\theequation d},\\
& {\kern 2em}{y_i} = {y_{n,}}\forall i,n \in {\cal K},{\rm mod}(i,N)={\rm mod}(n,N), \label{E-opti problemE} \tag{\theequation e}
\end{align}
where $\left\lfloor  \cdot  \right\rfloor $ is the floor function and $\bmod \left( {a,b} \right)$ denotes the remainder of dividing $a$ by $b$. The goal of problem (\ref{opti problem}) is to find the optimal placement of LEDs that minimizes power consumption, where (\ref{E-opti problemA}) and (\ref{E-opti problemB}) denote the illumination level and communication constraints, respectively. (\ref{E-opti problemC}) denotes the illumination uniformity constraint. In addition, (\ref{E-opti problemD}) and (\ref{E-opti problemE}) indicate that the LEDs are arranged in a rectangular array.

\section{Proposed LED Placement Algorithm}\label{proposed}
The problem (\ref{opti problem}) is non-convex due to the constraint (\ref{opti problem}a), (\ref{opti problem}b), and (\ref{opti problem}c). Hence its direct solution is computationally intractable. In this section, we first transform the illumination uniformity constraint (\ref{opti problem}c) into a series of linear constraints. Then, we further propose an iterative algorithm to transform the problems into a series of convex subproblems. Finally, the
Lagrange dual method is used to obtain the optimal solution.

\subsection{{\rm CV(RMSE)} Constraint Transformation}\label{CVtrans}
In this subsection, we transform the constraint (\ref{E-opti problemC}) into linear constraints to make (\ref{opti problem}) mathematical tractable.

As shown in Fig. \ref{scenario}, there are $M \times N$ LEDs installed. Let the distance between adjacent LEDs along $x$ and $y$ coordinates be $L_x$ and $L_y$, respectively. In addition, let the distance from the left/right/up/down wall to its nearest LED be $L_{lx}$/$L_{rx}$/$L_{uy}$/ $L_{dy}$. In this work, we assume that LEDs are placed symmetrically, i.e., $L_{lx}$=$L_{rx}$ and $L_{uy}$=$L_{dy}$. Note that, to the best of authors' knowledge, most of the existing studies investigated scenarios where LEDs are symmetrically placed \cite{Wang2012Performance,stefan2013analysis,liu2014cellular,hussein2015mobile,niaz2016deployment,yang2008uniform}. Therefore, this assumption will not impair applying this work to the typical VLC scenarios. Based on the symmetric assumption, we have:
\vspace{-0.2cm}
\begin{equation}\label{optLx}
\vspace{-0.2cm}
{L_{lx}} = {L_{rx}} = \frac{{{x_l} - (M - 1){L_x}}}{2},
\end{equation}
and
\begin{equation}\label{optLy}
\vspace{-0.2cm}
{L_{uy}} = {L_{dy}} = \frac{{{y_l} - (N - 1){L_y}}}{2}.
\end{equation}
From (\ref{optLx}) and (\ref{optLy}), the position of each LED $i$ can be expressed as:
\vspace{-0.2cm}
\begin{equation}\label{xi}
\vspace{-0.2cm}
{x_i} = \left\lfloor {\frac{i-1}{N}} \right\rfloor {L_x} + {L_{lx}},
\end{equation}
and
\begin{equation}\label{yi}
{y_i} = {\rm mod}(i-1,N) {L_y} + {L_{dy}},
\end{equation}

The the expression for CV(RMSE) can be derived from (\ref{cvrmse}) with respect to $L_x$ and $L_y$ as:
\vspace{-0.2cm}
\begin{equation}\label{CV-fxy}
\begin{array}{l}
\rm{CV\left( {RMSE} \right)}{\rm{ = }}\\
\!\!\!\frac{{\!\sqrt {\!{{\sum\limits_{j = 1}^U\!\! {\left(\! {\sum\limits_{i = 1}^K \!\! {{P_i}\!\!\left[\! \begin{array}{l}
\!\!U{\left(\! {{{\left( \!\!{{A_i}{L_x} \!+ \!{B_j}}\! \right)}\!^2} \!+\! {{\left(\! {{C_i}{L_y} \!+\! {D_j}} \! \right)}\!^2} + {h^2}} \right)\!\!^{ - \frac{{\left( {m + 3} \right)}}{2}}}- \!\!\sum\limits_{n = 1}^U\! {{{\left(\! {{{\left(\! \! {{A_i}{L_x}\! + \!{B_n}} \right)}^2} \!+\! {{\left(\! {{C_i}{L_y} \!+\! {D_n}} \!\right)}^2}\! + \!{h^2}} \!\right)}\!^{ - \frac{{\left( {m + 3}  \right)}}{2}}}}
\end{array}\! \!\right]} }\! \right)} \!}^2}\!} }}{{\sqrt U \sum\limits_{i = 1}^K {\sum\limits_{j = 1}^U {{P_i}{{\left( {{{\left( {{A_i}{L_x} + {B_j}} \right)}^2} + {{\left( {{C_i}{L_y} + {D_j}} \right)}^2} + {h^2}} \right)}^{ - \frac{{\left( {m + 3} \right)}}{2}}}} } }},
\end{array}
\end{equation}
where ${A_i} = \left\lfloor {\frac{i-1}{N}} \right\rfloor - \left( {M - 1} \right)/2$, ${B_j} = x_{l}/2 - {x_j}$, ${C_i} = {\rm mod}(i-1,N) - \left( {N - 1} \right)/2$, and ${D_j} = y_{l}/2 - {y_j}$. Therefore, for fixed LED power $P_i$, (\ref{CV-fxy}) becomes a function of $L_x$ and $L_y$.
When one of $L_x$ or $L_y$ fixed, the value of the other one that satisfies (\ref{E-opti problemC}) can be obtained by bisection method. Note that with the feasible ranges of $L_x$ or $L_y$, the feasible ranges of $x_i$ or $y_i$ can be obtained directly according to (\ref{xi}) and (\ref{yi}),  and thus the complex illumination uniformity constraint can be transformed into a series of linear constraints. From the above analysis, we can conclude that once the LED power and the x or y coordinate are given, the illumination uniform constraint (\ref{E-opti problemC}) can be transformed into linear constraints. The strategy of obtaining LED power and iteratively fixing x and y coordinates of LEDs will be specified in Section \ref{Loop}.

\subsection{Location optimization}\label{Loop}
In section \ref{CVtrans}, we analyzed the illumination uniformity constraint (\ref{E-opti problemC}). However, solving problem (\ref{opti problem}) should also consider the non-convex constraints of (\ref{E-opti problemA}) and (\ref{E-opti problemB}). In addition, the optimization process of each LED is also interdependent. To solve these challenges, an iterative algorithm is proposed. In the proposed iterative algorithm, (\ref{opti problem}) is decoupled into a series of interdependent subproblems. Each subproblem corresponds to the optimization of a current LED's x-coordinate or y-coordinate, and all the former solutions (i.e., the locations and the power of LEDs that have been optimized) of the sub-problems will be plugged into the current sub-problem. Then, the optimization problem of ${x_i}$ and ${y_i}$ of the current LED $i$ can be expressed as:
\begin{align}\label{E-opti subproblem1}
   \setlength{\abovedisplayskip}{1pt}
   \setlength{\belowdisplayskip}{1pt}
&\mathop {\min }\limits_{{x_i}} {P_i} \\
&{\rm s.t.}{\kern 1em}{P_i} \ge {I _{ij}}Vd_{ij}^{m + 3},\forall i \in {\cal K},\forall j \in {{\cal U}_i}\label{E-opti subproblem1A} \tag{\theequation a}\\
&{\kern 2em}{P_i} \ge {C_j}Wd_{ij}^{m + 3},\forall i \in {\cal K},\forall j \in {{\cal U}_i} \label{E-opti subproblem1B} \tag{\theequation b},\\
& {\kern 2em}  {x_i} \in {\cal R}_{x_i}, \forall i \in {\cal K}, \label{E-opti subproblem1C} \tag{\theequation c}\\
& {\kern 2em}{x_i} = {x_{n,}}\forall i,n \in {\cal K},\left\lfloor {i/N} \right\rfloor  = \left\lfloor {n/N}\right\rfloor, \label{E-opti subproblem1D}\tag{\theequation d}
\end{align}
and
\begin{align}\label{E-opti subproblem2}
   \setlength{\abovedisplayskip}{1pt}
   \setlength{\belowdisplayskip}{1pt}
&\mathop {\min }\limits_{{y_i}} {P_i} \\
&{\rm s.t.}{\kern 1em}{P_i} \ge {I _{ij}}Vd_{ij}^{m + 3},\forall i \in {\cal K},\forall j \in {{\cal U}_i}\label{E-opti subproblem12A} \tag{\theequation a}\\
&{\kern 2em}{P_i} \ge {C_j}Wd_{ij}^{m + 3},\forall i \in {\cal K},\forall j \in {{\cal U}_i} \label{E-opti subproblem12B} \tag{\theequation b},\\
& {\kern 2em} {y_i} \in {\cal R}_{y_i}, \forall i \in {\cal K}, \label{E-opti subproblem2c} \tag{\theequation c}\\
& {\kern 2em}{y_i} = {y_{n,}}\forall i,n \in {\cal K},{\rm mod}(i,N)={\rm mod}(n,N) ,\tag{\theequation d}
\end{align}
respectively, where $V=\frac{2\pi }{\xi (m+1)Ag\left( {{\psi }_{ij}} \right){{H}^{m+1}}}$, $W = \frac{{{{\left( {2\pi } \right)}^{3/2}}{e^{ - 1/2}}\sqrt {{\sigma _\omega }^2 + \sum\nolimits_{n = 1,n \ne i}^K {(\xi {P_n}{h_{nj}}} {)^2} } }}{{\xi (m + 1)Ag\left( {{\psi _{ij}}} \right){H^{m + 1}}}}$ and ${C_j}=\sqrt {{2^{2{c_{th,j}}}} - 1}$.  $V$ and $W$ are obtained by substituting (\ref{E-channel gain}) into (\ref{E-opti problemA}) and (\ref{E-opti problemB}), respectively. ${\cal R}_{x_i}$ and ${\cal R}_{y_i}$\footnote{ In the first iteration, the ranges of each ${x_i}$ and ${y_i}$ are set to ${x_i} \in \left[ {\left\lfloor {\frac{{i - 1}}{N}} \right\rfloor  \cdot \frac{{{x_l}}}{M},\left\lfloor {\frac{i}{N}} \right\rfloor  \cdot \frac{{{x_l}}}{M}} \right]$ and ${y_i} \in \left[ {\bmod \left( {i - 1,N} \right) \cdot \frac{{{y_l}}}{N},\bmod \left( {i,N} \right) \cdot \frac{{{y_l}}}{N}} \right]$, respectively. After the first iteration, the range ${\cal R}_{x_i}$ and ${\cal R}_{y_i}$ can be obtained according to Algorithm \ref{algorithm-all}.}
represent the linear ranges of ${x_i}$ and ${y_i}$ for illumination uniformity, respectively. Both of them may contain several subsets. We will first obtain all LEDs' ${x_i}$ by solving (\ref{E-opti subproblem1}) with given ${y_i}$, and ${y_i}$ can be updated from (\ref{E-opti subproblem2}) with obtained ${x_i}$. Then, ${x_i}$ can be updated again utilizing the obtained ${y_i}$. The iterations end until the sum power is not able to be further reduced, and after each iteration, the updated x or y coordinate and the power can be plugged into (\ref{CV-fxy}) to obtain the range of $L_x$ or $L_y$.

As we can observe, (\ref{E-opti subproblem1}) and (\ref{E-opti subproblem2}) have the same form. Therefore, we first solve (\ref{E-opti subproblem1}), and (\ref{E-opti subproblem2}) can be solved in a similar way. According to (\ref{E-opti subproblem1D}), we only need to optimize the x-coordinate of the first LED. Then, the distance ${L_x}$ can be obtained by $L_x = \left( {x_l - 2{x_1}} \right)/\left( {M - 1} \right)$ and the rest LEDs' x-coordinates can be obtained according to (\ref{xi}). Therefore, solving (\ref{E-opti subproblem1}) is equivalent to solving:
\begin{align}\label{E-cvx problem}
   \setlength{\abovedisplayskip}{1pt}
   \setlength{\belowdisplayskip}{1pt}
&\mathop {\min }\limits_{{x_i}} {P_i} \\
&{\rm s.t.}{\kern 1em}{P_i}^{\frac{2}{{m + 3}}} \ge {M_{ij}}d_{ij}^2, i = 1,\forall j\in{{\cal U}_i}, \label{E-cvx problemA} \tag{\theequation a}\\
& {\kern 2em}  {x_i} \in {\cal R}_{x_i}, i = 1, \label{E-cvx problemC} \tag{\theequation b}
\end{align}
for the first LED and solving:
\begin{align}\label{E-cvx problem1}
   \setlength{\abovedisplayskip}{1pt}
   \setlength{\belowdisplayskip}{1pt}
& \min {P_i}{\kern 16em}\\
& {\rm{s.t.}}{\kern 1em}{P_i}^{\frac{2}{{m + 3}}} \ge {M_{ij}}d_{ij}^2,\forall i \in {\cal K},i \ne 1,\forall j\in{{\cal U}_i},\label{E-cvx problem1A} \tag{\theequation a}
\end{align}
for the rest LEDs, where:
   \setlength{\abovedisplayskip}{1pt}
   \setlength{\belowdisplayskip}{1pt}
\begin{equation}\label{Mj}
 {{M}_{ij}}={{\left(\max\left\{{{I}_{ij}}V,{{C}_{j}}W\right\}\right)}^{\frac{2}{m+3}}}.
\end{equation}
From (\ref{E-cvx problem}) and (\ref{E-cvx problem1}), we can see that the value of the power is exactly determined by ${{M}_{ij}}$ which is related to $V$, $W$, and the data rate and illumination level constraints of the receivers. 
In each sub-problem (\ref{E-cvx problem}) and (\ref{E-cvx problem1}), ${{M}_{ij}}$ can be regraded as a constant since the former solutions that optimize the locations of the former LEDs have been updated.

It is easy to prove that both of objective function and constraint in (\ref{E-cvx problem}) are convex. Therefore, Lagrangian dual method can be used to solve the problem. We use $\bm{\lambda}  = {\left[ {{\lambda _1},{\lambda _2},...,{\lambda _U}} \right]^T}, \lambda_j>0, j=1,...U$ to denote the dual variable associated with each constraint $j$ in (\ref{E-cvx problemA}). We represent the dual problem (\ref{E-cvx problem}) as:
\begin{equation}\label{dual}
   \setlength{\abovedisplayskip}{1pt}
   \setlength{\belowdisplayskip}{1pt}
\mathop {\max }\limits_{\lambda \ge 0} g(\bm{\lambda} ),
\end{equation}
where
\begin{align}\label{lang}
& g(\bm{\lambda} ) = \mathop {\min }\limits_{{x_i}} {\cal L}({x_i},{P_i},\bm{\lambda} ),\\
&{\rm s.t.}{\kern 1em}{x_i} \in {\cal R}_{x_i}, i = 1.\label{langA} \tag{\theequation a}
\end{align}
The Lagrange function in (\ref{lang}) can be expressed as:
\begin{equation}\label{E-lagrange}
{\cal L} \!= {P_i}+\sum\nolimits_{j\in {{\cal U}_i}}{{\lambda _j}} ({M_{ij}}({({x_i}- {x_j})\!^2}+ {({y_i}- {y_j})^2}+{H^2})- {P_i}\!^{\frac{2}{{m + 3}}}).
\end{equation}

Based on Karush Kuhn-Tucker (KKT) conditions, by taking the first derivative of (\ref{E-lagrange}) with respect to ${{P}_{i}}$ and ${{x}_{i}}$, we will have:
\begin{equation}\label{derivative-P}
\setlength{\abovedisplayskip}{1pt}
\setlength{\belowdisplayskip}{1pt}
\frac{{\partial {\cal L}}}{{\partial {P_i}}} = 1 - \frac{2}{{m + 3}}\sum\nolimits_{j \in {U_i}} {{\lambda _j}} {P_i}^{ - \frac{{m + 1}}{{m + 3}}} = 0,
\end{equation}
\begin{equation}\label{derivative-x}
\frac{{\partial {\cal L}}}{{\partial {x_i}}} = 2\sum\nolimits_{j \in {U_i}} {{\lambda _j}} {M_{ij}}({x_i} - {x_j}) = 0,
\end{equation}
By solving (\ref{derivative-P}) and (\ref{derivative-x}), we have:
 \begin{equation}\label{E-cvx Pi}
\setlength{\abovedisplayskip}{2pt}
\setlength{\belowdisplayskip}{2pt}
{P_i} = {\left( {\frac{2}{{m + 3}}\sum\nolimits_{j \in {{\cal U}_i}} {{\lambda _j}} } \right)^{\frac{{m + 3}}{{m + 1}}}},
\end{equation}

\begin{equation}\label{E-cvx_xy}
\setlength{\abovedisplayskip}{2pt}
\setlength{\belowdisplayskip}{2pt}
{x_i} = \frac{{\sum\nolimits_{j \in {{\cal U}_i}} {{\lambda _j}} {M_{ij}}{x_j}}}{{\sum\nolimits_{j \in {{\cal U}_i}} {{\lambda _j}} {M_{ij}}}}.
\end{equation}
The Lagrange function in (\ref{lang}) is a quadratic function with respect to $x_i$. Therefore, the optimal $x_i$ that minimizes $\cal L$ can be obtained as:
\begin{equation}\label{qfconstraint}
x_i^* = \left\{ \begin{array}{l}
\arg \mathop {\min }\limits_{{x_i} \in {\cal A}} L({x_i},{y_i},{P_i},\lambda ), \text{if } {{\hat x}_i} \notin {{\cal R}_{{x_i}}}\\
{{\hat x}_i}, \text{if } {{\hat x}_i} \in {{\cal R}_{{x_i}}}
\end{array} \right.
\end{equation}
where ${{\hat x}_i}=\frac{{\sum\nolimits_{j \in {U_i}} {{\lambda _j}} {M_{ij}}{x_j}}}{{\sum\nolimits_{j \in {U_i}} {{\lambda _j}} {M_{ij}}}}$ and ${\cal A}$ consists of all endpoints of ${\cal R}_{x_i}$. For instance, if ${\cal R}_{x_i}$ consists of set $[1,3]$ and set $[5,8]$, we have ${\cal A} = \{1,3,5,8\}$, and we set the optimal $x_i$ only be equal to 1, 3, 5, or 8 if ${{\hat x}_i} \notin {R_{{x_i}}}$.

Given ${{P}_{i}}$ and ${{x}_{i}}$, we use a subgradient method \cite{boyd2004convex} to update the value of ${\lambda _j}$. The updating procedure of ${\lambda _j}$ is given by:
\begin{equation}\label{E-cvx lmd}
\lambda _j^{(l + 1)} = \lambda _j^{(l)} + \gamma ({M_{ij}}({({x_i}- {x_j})^2}+ {({y_i}- {y_j})^2}+{H^2})- {P_i}^{\frac{2}{{m + 3}}}),
\end{equation}
where $\gamma$ is a dynamic step-size of iteration $l$ $(l \in {1,2,...,L_i})$ and $L_i$ is the maximum number of iterations for optimizing the LED $i$. When optimizing the LED other than the first one, the position of the LED can be easily obtained with (\ref{xi}), and the optimal power of the LED can be obtained by using (\ref{E-cvx Pi}) and (\ref{E-cvx lmd}) only. The iterative algorithm for LEDs' x-coordinate optimization (LXO) is summarized in \textbf{Algorithm} \ref{algorithmx}.
\begin{algorithm}[htpb]
  \caption{LEDs' x-coordinate Optimization (LXO) Algorithm}\label{algorithmx}
    \textbf{Initialize:} Lagrange multipliers ${{\lambda }_{j}}$ and all LEDs' y-coordinates $y_i, \forall i \in {\cal K}$.
    \begin{algorithmic}[1]
    \Repeat
    \For {$i=1$ to $K$}
      \State Update ${{M_{ij}}}$ according to (\ref{Mj}).
      \State Update power ${{P_i}}$ based on (\ref{E-cvx Pi}).
       \State Update ${x_i}$ according to (\ref{E-cvx_xy})-(\ref{qfconstraint}) if $i=1$, and ${x_i} = \left\lfloor {\frac{i}{N}} \right\rfloor {L_x} + \frac{{{x_l} - (M - 1){L_x}}}{2}$ if $i \ne 1$.
      \State Update Lagrange multipliers ${\lambda _j}, j \in {{\cal U}_i},i \in {\cal K}$ according to (\ref{E-cvx lmd}).
    \EndFor
    \Until   $\sum\nolimits_{i \in {\cal K}} {{P_i}}$ converges.
    \Ensure  Power ${{P_i}}$ and location of each LED.   
  \end{algorithmic}
\end{algorithm}

The subproblem (\ref{E-opti subproblem2}) is similar to (\ref{E-opti subproblem1}). Hence, we use the same algorithm to optimize the y-coordinates of LEDs, which is omitted for simplification here. We define the algorithm of solving (\ref{E-opti subproblem2}) as LEDs' y-coordinate Optimization (LYO) Algorithm. The obtained ${x_i}$ and ${y_i}$ in (\ref{E-opti subproblem1}) and (\ref{E-opti subproblem2}) can be further iteratively optimized, until the power and the locations of all LEDs converge.

The complete LED placement optimization algorithm (LXYU) that iteratively conducts LXO and LYO with the illumination uniformity constraint is summarized in \textbf{Algorithm} \ref{algorithm-all}. We first use LXO and LYO to obtain x-coordinate and y-coordinate of LEDs, respectively. Then, we use bisection method to calculate the boundaries of each LED's coordinate from the function CV(RMSE). By iterating LXO and LYO with CV(RMSE) function respectively, we can obtain the solution of ${x_i}$ or ${y_i}$. 
The power and the locations of all LEDs will finally converge to the optimal solutions.

\begin{algorithm}[htpb]
  \begin{algorithmic}[1]
  \caption{Complete Algorithm for LED Placement Optimization (LXYU)}\label{algorithm-all}
    \Require Locations, data rate constraint $c_{th}$ and illumination level constraint $\eta_{th}$ of receivers in ${\cal U}$. The height $H$ and y-coordinate of LEDs.\\
    \Comment {\kern 0.2em} Cell-association. Ranges of each LED' coordinate.
   \Repeat
     \Repeat
       \State Update ${{P_i}},\forall i \in {\cal K}$ and ${x_i}, i=1$ calculated by \textbf{LXO Algorithm}.
        \State Calculate ranges of $L_x$ with $U_{th}$ and ${{P_i}},i \in {\cal K}$ according to (\ref{cvrmse}) and (\ref{E-opti problemC}).
         \State Calculate the ranges ${\cal R}_{xi}$ of LEDs' x-coordinates with the range of $L_x$.
     \Until ${\rm{CV}(RMSE)} \le U_{th}$.
     \Repeat
       \State Update ${{P_i}},\forall i \in {\cal K}$ and ${y_i}, i=1$ calculated by \textbf{LYO Algorithm}.
        \State Calculate ranges of $L_y$ with $U_{th}$ and ${{P_i}},i \in {\cal K}$ according to (\ref{cvrmse}) and (\ref{E-opti problemC}).
        \State Calculate the ranges ${\cal R}_{yi}$ of LEDs' y-coordinates with the range of $L_y$.
     \Until ${\rm{CV}(RMSE)} \le U_{th}$.
    \Until $\sum\nolimits_{i \in {\cal K}} {{P_i}}$ converges.
    \Ensure  $P{\rm{ = }}\sum\nolimits_{i \in {\cal K}} {{P_i}}$.   
  \end{algorithmic}
\end{algorithm}
\subsection{Complexity Analysis}\label{complexity}
We assume that the maximum number of the iterations for sub-problems (\ref{E-opti subproblem1}) and (\ref{E-opti subproblem2}) is $S_1$. Moreover, $S_2$ is the maximum number of the iterations for LXO and CV(RMSE) function, while $S_3$ is the maximum number of the iterations for LYO and CV(RMSE) function. The complexity of LXO algorithm is ${\cal O}(\max \left\{ {{L_1},{L_2},...,{L_K}} \right\}\sum\nolimits_{i = 1}^K {\left| {{{\cal U}_i} } \right|})$ and that of LYO algorithm is ${\cal O}(\max \left\{ {{L'_1},{L'_2},...,{L'_K}} \right\}\sum\nolimits_{i = 1}^K {\left| {{{\cal U}_i} } \right|})$, where $L_i$ and $L'_i$ denote the  maximum number of iterations for updating the power of LED $i$ in LXO and LYO algorithms, respectively. $\left|{{\cal U}_i}\right| $ is the number of receivers associated with LED $i$. In addition, when solving $L_x$ and $L_y$ in CV(RMSE) function, the complexity is $ {\cal O}({\log _2}\frac{{{x_l}}}{{\left( {M - 1} \right)\varepsilon }})$ and $ {\cal O}({\log _2}\frac{{{y_l}}}{{\left( {N - 1} \right)\varepsilon }})$ \cite{kelley1995iterativebook}, respectively, where $\varepsilon$ represent the tolerance in bisection method. Thus, the complexity of \textbf{Algorithm} \ref{algorithm-all} can be calculated as $ {\cal O}({S_1}\max \left\{  Q_1,Q_2\right\}\sum\nolimits_{i = 1}^K {\left| {{{\cal U}_i}} \right|} )$, where $Q_1={S_2} \max \left\{ {{L_1},{L_2},...,{L_K},{\log _2}\frac{{{x_l}}}{{\left( {M - 1} \right)\varepsilon }}} \right\}$ and $Q_2 = {S_3} \max \left\{ {{L'_1},{L'_2},...,{L'_K},{\log _2}\frac{{{y_l}}}{{\left( {N - 1} \right)\varepsilon }}} \right\}$. On the other hand, the complexity of exhaustive search is ${\cal O}({T^K})$, where $T$ is the total number of LEDs' possible positions on the ceil. If an accurate LED placement is conducted, the value of $T$ can be extremely large. For instance, for a  7.5 m $\times$ 5 m $\times$ 3 m room with 4 LEDs, $T=3750$ when using 0.1 m search step, and the search complexity is ${\cal O}(  1.9775*10^{14})$. Thus, the complexity of exhaustive search is prohibitive in this case.

\section{Numerical Results}\label{results}
We consider a rectangular 7.5 m $\times$ 5 m $\times$ 3 m room, which is further equally divided into $K$ sub-areas. In each sub-area, an LED is deployed to provide communication and illumination service. We set 160 receivers at the receiver plane, which are uniformly distributed. The system parameters are listed in Table \ref{parameters}.

\begin{table}[htbp]
		\centering \caption{System parameters}\label{parameters}
            \footnotesize
		\begin{tabular}{l|c}\hline
			\multicolumn{1}{c|} {Parameters}&Values\\\hline
			Semi-angle at half power ${\Phi _{1/2}}$&$60^\circ$ \\\hline
            Receiver FOVs' semi-angle ${\Psi _c}$&$60^\circ$ \\\hline
            Detect area of photodiodes $A$&1 cm$^2$ \\\hline
            Refractive index ${n_r}$&1.5 \\\hline
            Dynamic step-size ${\gamma_j}$&0.01 \\\hline
            SNR&20 dB \\\hline
		\end{tabular}
\end{table}

For comparison, we consider two baseline schemes in which the illumination uniformity is not considered. First, LED-centered algorithm (LCA) is implemented, in which each LED is at the center of each sub-area. We also simulate the placement that is optimized by LXO and LYO Algorithm only (LXYO).
\begin{figure}[htbp]
\vspace{-0.1cm}
\centering
\subfigure[Illumination distribution of LCA.]{\includegraphics[height=4.5cm,trim=30 0 280 10,clip]{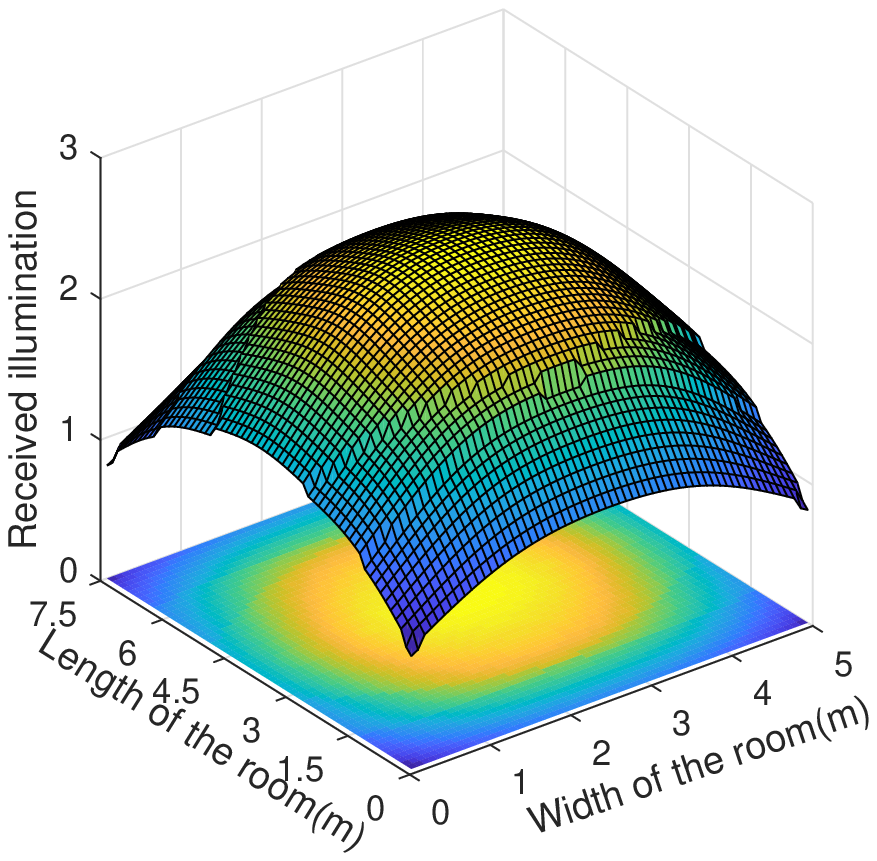}}
\subfigure[Illumination distribution of LXYU.]{\includegraphics[height=4.5cm,trim=280 0 40 10,clip]{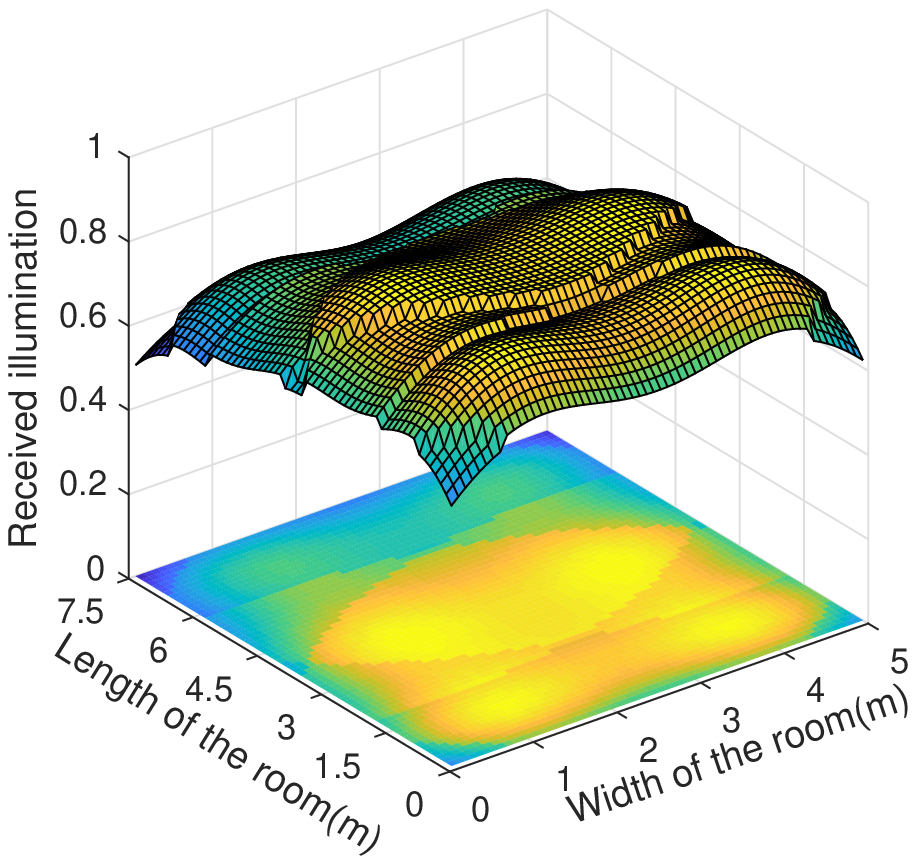}}
\vspace{-0.1cm}
\caption{ Received illumination distribution of LCA and LXYU with 6 LEDs. }\label{compare}
\vspace{-0.1cm}
\end{figure}

In Fig. \ref{compare}, we compare the received illumination distribution of LCA and LXYU with 6 LEDs. We set the uniformity tolerance, data rate constraint, and the illumination level constraint as 0.1, 0.8, and 0.4, respectively. From Fig. \ref{compare}, we can see that LXYU can achieve more uniform illumination distribution, and the CV(RMSE) is reduced from 0.22 to 0.1.
\begin{figure}[htbp]
\vspace{-0.1cm}
\centering
\includegraphics[height=5cm,trim=10 5 20 15,clip]{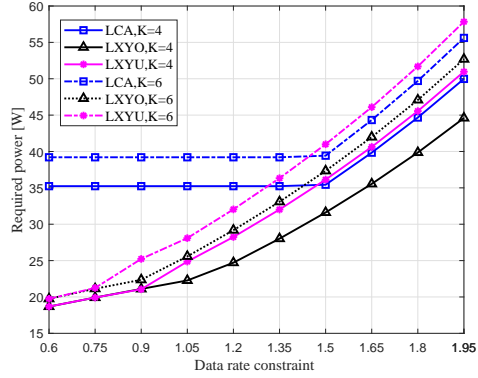}
\caption{Sum power consumption of LEDs versus the data rate constraint.}\label{Cth}
\vspace{-0.1cm}
\end{figure}

Figure \ref{Cth} shows how the sum power consumption changes as the data rate constraint of the receivers varies when 4 and 6 LEDs are placed. The illumination uniformity constraint and the illumination level constraint are set to 0.16 and 0.4, respectively. From Fig. \ref{Cth}, the sum power increases with the data rate constraints. It can also be observed that the proposed algorithm LXYU is efficient. For instance, when there are 4 LEDs and the data rate constraint is 1.05, LXYU can reduce $22.86\%$ power consumption compared to LCA. However, LXYU consumes more power than LXYO, since it takes the illumination uniformity constraint into consideration. In addition, when the data rate constraint is less than 1.5 bit/transmission, the sum power of LCA is a constant. The reason is that in this circumstance, the power required for illumination is more than that required for communication, and thus the power is determined by illumination level constraint.
\begin{figure}[htbp]
\vspace{-0.1cm}
\centering
\includegraphics[height=5cm,trim=10 5 20 15,clip]{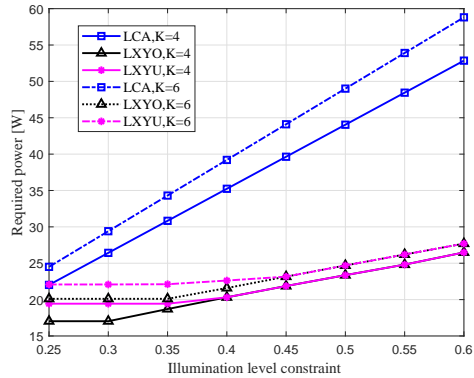}
\caption{Sum power consumption of LEDs versus the illumination level constraint.}\label{etath}
\vspace{-0.2cm}
\end{figure}
\begin{figure}[htbp]
\vspace{-0.1cm}
\centering
\includegraphics[height=5cm,trim=10 5 20 15,clip]{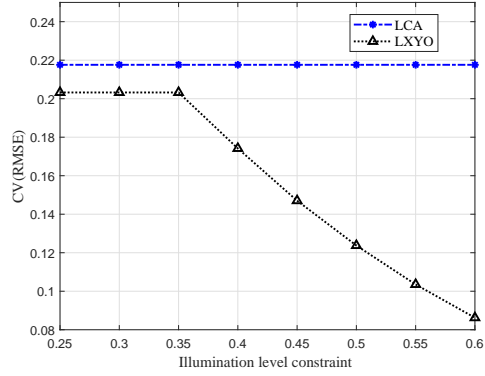}
\caption{The illumination uniformity versus the illumination level constraint.}\label{figcvrmse}
\vspace{-0.1cm}
\end{figure}
\begin{figure}[htbp]
\vspace{-0.1cm}
\centering
\includegraphics[height=5cm,trim=10 0 20 23,clip]{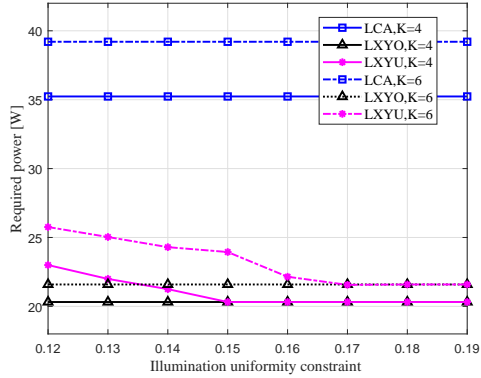}
\caption{Sum power consumption of LEDs versus the illumination uniformity constraint.}\label{Uth}
\vspace{-0.3cm}
\end{figure}

Figure \ref{etath} shows the sum power consumption as a function of the illumination level constraints of the receivers. Two cases with 4 and 6 LEDs are considered. The data rate constraint is set to 0.8 and the illumination uniformity constraint is set to 0.16. From Fig. \ref{etath}, we can observe that LXYU can achieve less power consumption, and compared to LCA, the reduced power of LXYU increases as illumination level constraint increases. When there are 6 LEDs and the illumination level constraint is 0.45, LXYU can reduce 47.48$\%$ power consumption compared to LCA. It can also be observed that the power consumptions of LXYO and LXYU are the same when the illumination level constraint is more than 0.45. This is because in the considered configuration, the illumination uniformity increases with the illumination level constraint, as will be verified in Fig. \ref{figcvrmse}. Therefore, when the illumination level constraint is high enough, LXYO can naturally satisfy the illumination uniformity constraint even though it is not considered in LXYO. In this way, LXYU and LXYO have the same power consumption when the illumination level is higher than 0.45.

Figure \ref{figcvrmse} shows the illumination uniformity of LCA and LXYO with the increase of illumination level constraints. The data rate constraint is set to 0.8 and the illumination uniformity constraint is set to 0.16. The illumination uniformity constraint is quantified by CV(RMSE). In particular, the higher the illumination uniformity constraint is, the worse the illumination uniformity performance is. Fig. \ref{figcvrmse} further confirms that the illumination uniformity of LXYO gets better as the illumination level constraint increases. That indicates that when the illumination level constraint reaches a certain value, the CV(RMSE) of LXYO is already less than the set illumination uniformity constraint. That is why the power consumptions of LXYO and LXYU are the same when the illumination level constraint is more than 0.45 in Fig. \ref{etath}.

In Fig. \ref{Uth}, we analyze the required sum power for different illumination uniformity constraints. The data rate constraint and the illumination level constraint are set as 0.8 and 0.4, respectively. From Fig. \ref{Uth}, it can be observed that when compared with LCA, LXYO can reduce the power consumption by 43.32$\%$ when placing 4 LEDs, and 44.92$\%$ when placing 6 LEDs. We can also observe that the sum power of LXYU decreases as the constraint of illumination uniformity increases. When the illumination uniformity threshold $U_{th}$ is large enough, LXYU has the same power consumption with LXYO. This is because in such cases LXYO can naturally satisfy the illumination uniformity constraint even without considering it. In this way, LXYO and LXYU have the same performance.

\vspace{-0.cm}
\section{Conclusion}\label{Conclusion}
In this paper, we have studied the problem of optimal LED placement that takes both the illumination and communication interactions of LEDs into consideration. We have formulated the problem as a power minimization problem under both communication and illumination level constraints. Since the formulated problem is non-convex and the optimization variables are interdependent, we have transformed the illumination uniformity constraint into a series of linear constraints, and then proposed an iterative algorithm to solve the problem. The proposed iterative algorithm transformed the original problem into a series of convex sub-problem which were efficiently solved by Lagrange dual method.  Numerical results show that the proposed approach can yield more than $22.86\%$ improvements in power efficiency.

\section*{Funding}
This work was supported by  National Natural Science Foundation of China (61871047), National Natural Science Foundation of China (61901047), Beijing Natural Science Foundation (4204106) and China Postdoctoral Science Foundation (2018M641278).

\bibliography{Citation}
\end{document}